\documentclass[12pt]{article}
\usepackage{latexsym}\usepackage{epsfig,amssymb,euscript}
\usepackage{amsmath} \usepackage{bbm,slashed}
\usepackage[usenames,dvipsnames]{xcolor}
\usepackage{ulem}

\newcommand{\Lorenzo}[1]{{\color{OliveGreen} #1}}
\newcommand{\be}{\begin{equation}}
\newcommand{\ee}{\end{equation}}
\newcommand{\bea}{\begin{eqnarray}}
\newcommand{\eea}{\end{eqnarray}}
\newcommand{\tr}{\mathrm{tr\,}}

\begin{document}
\begin{titlepage}
\begin{flushright}
\end{flushright}
\bigskip
\def\thefootnote{\fnsymbol{footnote}}

\begin{center}
\vskip -10pt
{\LARGE
{\bf
Holographic Correlators for \\
\vspace{0.3in}
General Gauge Mediation
}
}
\end{center}

\bigskip
\begin{center}
{\large
Riccardo Argurio$^{1,2}$
Matteo Bertolini$^{3,4}$,
Lorenzo Di Pietro$^3$, \\
\vskip 5pt
Flavio Porri$^3$ and
Diego Redigolo$^{1,2}$}

\end{center}

\renewcommand{\thefootnote}{\arabic{footnote}}

\begin{center}
$^1${Physique Th\'eorique et Math\'ematique \\
Universit\'e Libre de Bruxelles, C.P. 231, 1050 Bruxelles, Belgium\\}
$^2${International Solvay Institutes, Brussels, Belgium}\\
$^3$ {SISSA and INFN - Sezione di Trieste\\
Via Bonomea 265; I 34136 Trieste, Italy\\}
$^4$ {International Centre for Theoretical Physics (ICTP)\\
Strada Costiera 11; I 34014 Trieste, Italy}


\end{center}

\noindent
\begin{center} {\bf Abstract} \end{center}
We use holographic techniques to compute two-point functions of operators belonging to a conserved current supermultiplet in theories which break supersymmetry
at strong coupling. These are the relevant quantities one has to compute in models of gauge mediation to determine the soft spectrum in
supersymmetric extensions of the Standard Model (SSM).  Such holographic approach can be used for diverse gravitational backgrounds, but here we focus, for definiteness,  on asymptotically AdS backgrounds. After presenting the general framework, we apply our formulas to two explicit examples  which differ by the nature of the SSM gauginos, which have Dirac or
Majorana masses, corresponding to models that respectively preserve or break R-symmetry. \vspace{1.6 cm}
\vfill

\end{titlepage}

\setcounter{footnote}{0}

\section{Introduction and motivation}

The AdS/CFT duality \cite{Maldacena:1997re,Witten:1998qj,Gubser:1998bc} has proven to be an incredibly powerful tool to study properties of strongly coupled gauge theories in terms of dual
weakly coupled gravitational backgrounds. The primary objects  the basic dictionary allows one to compute are correlators of gauge invariant
operators.

Our aim in this work is to apply holographic techniques (i.e. holographic renormalization \cite{Bianchi:2001de,Bianchi:2001kw,Skenderis:2002wp}) to compute two-point functions of gauge invariant operators
related by SUSY transformations in a strongly coupled theory, and see how these
correlators differ when SUSY is broken at low energies, i.e.~either
spontaneously or softly. The leading application we have in mind is
General Gauge Mediation \cite{Meade:2008wd} (GGM), where two-point functions
of operators belonging to a conserved current supermultiplet encode
all the information that is needed to derive the soft spectrum of
superpartners.

In this work, we will focus our attention on asymptotically AdS (AAdS) spacetimes, which correspond to dual theories with a non-trivial superconformal
fixed point in the UV.\footnote{Strictly speaking, theories which are
superconformal cannot break SUSY spontaneously, because $\langle
T_{\mu\nu}\rangle = \epsilon \eta_{\mu\nu}$ contradicts
$T^\mu_\mu=0$. The set ups we consider will be thought of as toy
models where the superconformal symmetry is broken explicitely, though
often in a strongly coupled way.}
The operators in the current supermultiplet are dual
to bulk fields belonging to a vector multiplet of a ${\cal N}=2$ gauged
supergravity in 5 dimensions. Our task will be to compute two-point
correlators from the linear fluctuations of such fields over a
background that breaks SUSY.

An interesting class of backgrounds can be obtained by
considering an effective gauged supergravity Lagrangian describing the interaction of the ${\cal N}=2$ gravity multiplet with one
hypermultiplet. The latter, known as the universal hypermultiplet, contains the dilaton, and another scalar
related to a squashing mode,
which has $m^2=-3$ (in AdS units) and is charged under the
R-symmetry. We will show that turning on the charged scalar is a
necessary condition in order to get, eventually, Majorana masses for the Supersymmetric Standard Model (SSM)  gauginos. On the contrary, pure dilatonic
backgrounds, which as such preserve the R-symmetry,
can (and do, in general) provide a Dirac mass for gauginos. The rest of the paper is organized as follows.

In section 2 we first briefly review the GGM formalism. Then, we discuss
the holographic recipe to compute GGM correlators by means of the renormalized action for a 5d massless vector multiplet, the bulk multiplet  dual to the current superfield. In doing so, we explain in detail how to use holographic renormalization for such fields (for previous works using a similar philosophy see \cite{McGuirk:2011yg,STun}).

In section 3 we present a 5d ${\cal N}=2$ gauged supergravity model which is the simplest possible model containing all necessary ingredients to let us treat several qualitatively different examples, all arising as consistent solutions of the same 5d equations of motion.

As a warm-up, we start in section 4 by considering a  pure (and hence supersymmetric) AdS background. The holographic GGM functions are those of a supersymmetric field theory. We use the holographic computation to check the validity of our approach, and as useful reference for the more interesting examples we consider  afterwards.

In section 5 we turn to a dilaton-domain wall background, originally found in \cite{Gubser:1999pk,Kehagias:1999tr} (see also \cite{Constable:1999ch}) in the context of full fledged type IIB supergravity, used at that time as a candidate gravitational dual for confining theories. The background breaks all supersymmetry but preserves the R-symmetry. We find non-vanishing values for the sfermion masses while, consistently, SSM gauginos do not acquire a (Majorana) mass term. Interestingly, we find a pole at zero momentum for one  fermionic correlator, which signals the presence of a Dirac-like (hence R-symmetry preserving) mass for the gauginos \cite{Buican:2009vv,Intriligator:2010be}.  Notice that in our strongly coupled theory, such Dirac mass contribution arises as a consequence of strong dynamics in the hidden sector.

Finally, in section 6, we turn-on a small R-symmetry breaking scalar profile, that we treat linearly and without backreaction on the dilaton-domain wall background. The linear approximation is sufficient to show how things change significantly. Indeed, we show that in this case a Majorana mass for the gauginos is generated. Moreover, the pole at zero momentum of the fermionic correlator, responsible for a Dirac-like contribution to gaugino masses in the pure dilatonic background,  automatically disappears, in remarkable agreement with field theory expectations \cite{Buican:2009vv,Intriligator:2010be}.

We conclude in section 7 with a summary of our results and an outlook on possible further applications.


\section{Holographic renormalization of a vector multiplet}
In general gauge mediation  \cite{Meade:2008wd} the basic objects one has to compute are correlators of operators belonging to the hidden sector current supermultiplet. The 5d gravity dual to such current superfield is a  ${\cal N}=2$ massless vector multiplet.  In this section we review why these claims are correct, write down the
correlators we will be interested in, and then provide the holographic recipe to compute them.

\subsection{Correlators for General Gauge Mediation}
Any gauge mediation model can be visualized as a supersymmetric Standard Model visible sector whose gauge degrees of freedom couple to the hidden sector as
\be
\int d^4x d^4\theta  g V {\cal J},
\ee
at leading order in the SSM gauge coupling constant $g$. In the above formula $V$ is a SSM vector multiplet (here and below we will suppress
indices and effectively assume a $U(1)$ SSM gauge group, for simplicity), and ${\cal J}$
is a   $\mathcal{N} = 1$ multiplet of a conserved current.

In 4d, such a multiplet
contains a scalar operator $J$ of conformal
dimension $\Delta_0=2$, a fermionic operator $j_{\alpha}$ of conformal dimension
$\Delta_{1/2}=5/2$ and a vector operator $j_{i}$ of conformal dimension
$\Delta_1=3$.

The current multiplet is associated to a global symmetry
of the hidden sector that one has then to weakly gauge and identify with the
visible sector gauge group. The soft spectrum at low energies is
completely determined by two-point functions between the hidden sector
currents, which in Euclidean momentum-space read
\begin{align}
&\langle J(k)J(-k)\rangle=C_{0}(k^{2})\ , \label{GGM0}\\
&\langle j_{\alpha}(k)\bar{j}_{\dot\alpha}(-k)\rangle= - \sigma_{\alpha\dot\alpha}^{i}k_{i}C_{1/2}(k^{2})\ ,\label{GGM1/2}\\
&\langle j_{i}(k)j_{j}(-k)\rangle=(k_{i}k_{j}-\eta_{ij}k^{2})C_{1}(k^2)\ ,\label{GGM1}\\
&\langle j_{\alpha}(k)j_{\beta}(-k)\rangle=\epsilon_{\alpha\beta}B_{1/2}(k^{2})\ ,\label{GGMB}\
\end{align}
where a factor of $(2\pi)^4 \delta^{(4)}(0)$ is understood. The $C_{s}$ functions are real and dimensionless while $B_{1/2}$
is in general complex, breaks R-symmetry, and has the dimension of a mass.
Unbroken supersymmetry dictates
\be
C_{0}(k^{2})=C_{1/2}(k^{2})=C_{1}(k^2)\ , \qquad\qquad B_{1/2}(k^{2})=0\ .
\label{susycandb}
\ee
Expanding the effective Lagrangian of the gauge supermultiplet in the gauge coupling $g$, one easily sees that
the soft masses for sfermions and gauginos can be expressed in terms of above correlators as
\begin{align}
&m^{2}_{\tilde{f}}=-g^{4}\int \frac{d^{4}k}{(2\pi)^{4}}\frac{1}{k^2}\left(3C_{1}(k^2)-4C_{1/2}(k^2)+C_{0}(k^2)\right)\ ,\label{sfermionmass}\\
&m_{\lambda}=g^2 B_{1/2}(0)\ . \label{gauginomass}
\end{align}
In general, one might like to consider hidden sectors which break
supersymmetry at strong coupling, and this makes the problem
of computing the correlators of the hidden sector currents a
difficult one from the field theoretical point of view.

Our purpose is then to provide the correct recipe to compute GGM correlators (\ref{GGM0})-(\ref{GGMB}) using the AdS/CFT
correspondence.  In what follows we discuss the various steps one should pursue to reach this goal.

First, one should
consider a $5d$ supergravity background being possibly a stable solution of (a
consistent truncation of) $10d$ Type IIB supergravity compactified
on some $5d$ internal manifold. This background represents, holographically,
our strongly coupled hidden sector. Since this
background is the source of supersymmetry breaking, we will focus
on solutions which do not preserve any supersymmetry.
For the time being, we will not need to specify any detail of the background, except
that we take it to be asymptotically AdS, that is
\be
ds^2 \underset{\overset{z\to0}{}}{\simeq} \frac{1}{z^2}\left( dz^2
+  (dx^i)^2 \right)~ .
\ee
On such background we should then study the fluctuations for the fields dual
to the conserved $U(1)$ current multiplet ${\cal J}$. According to the AdS/CFT field/operator correspondence, the 5d supergravity
multiplet dual to a current supermultiplet  ${\cal J}$ is a ${\cal N}=2$ massless vector multiplet $(D,\,\lambda,\,A_\mu)$ , as detailed in Table  \ref{N = 1current} (for asymptotically AdS
backgrounds, the scaling at the boundary is directly related to the
mass of supergravity fields).
\begin{table}
\begin{center}
\begin{tabular}{|c|c|c|c|c|}
\hline
 $4d$ op. & $\Delta$& $5d$  field & scaling & $AdS$ mass\\ \hline
 $J(x)$&$\Delta_0=2$ & $D(z,x)$& $D(z,x)\simeq z^{2}lnz \,d_{0}(x)$& $m^2_D = \Delta_0(\Delta_0-4)=-4$\\ \hline
 $j_{\alpha}(x)$&$\Delta_{1/2}=5/2$&$\lambda(z,x)$&$\lambda(z,x)\simeq z^{3/2}\lambda_{0}(x)$&$\vert m_{\lambda}\vert=\Delta_{1/2} - 2 = 1/2$\\ \hline
 $j_{i}(x)$&$\Delta_1=3$& $A_{\mu}(z,x)$& $A_{\mu}(z,x)\simeq a_{0\mu}(x)$& $m_{A}= (\Delta_1-2)^2-1=0$\\ \hline
  \end{tabular}
  \caption{$4d$ $\mathcal{N} =1$ current multiplet and dual supergravity fields}\label{N = 1current}
  \end{center}
  \end{table}

On general grounds the existence of such a multiplet is ensured
as long as the internal manifold has some isometries. In the context
of the full $10d$ theory it would be a difficult task to identify the
correct $10d$ fluctuations corresponding to the $5d$ vector multiplet
\cite{Skenderis:2006uy}.\footnote{At least for the standard case of
  $\text{AdS}_{5}\times S^{5}$ the problem has been fully solved in
  \cite{Kim:1985ez, Gunaydin:1984fk}.} For this reason we will
consider consistent truncations of the full theory to $5d$ gauged
supergravity. Notice, though, that since we want to compute two-point functions of the dual
operators, we need to consider the action for the vector multiplet
only to quadratic order in the fields (but of course to all order
in the background fields).

Given a $5d$ supergravity action, we have to apply the
holographic renormalization procedure \cite{Bianchi:2001de,
  Bianchi:2001kw,Skenderis:2002wp} in order to obtain a finite well
defined on-shell boundary action. At the end of this procedure we will
get a renormalized action $S_{\text{ren}}[d_{0},\lambda_0,a_{0}]$
quadratic in the sources. The two point functions are then computed
deriving twice with respect to the sources.

We recall that the leading boundary mode for the scalar, $d_{0}$, can
be directly identified with the source for the operator $J$. Similarly
for the vector, with the gauge choice $A_{z}=0$, the leading boundary
mode $a_{0i}$ is the source for the $4d$ conserved vector
current at the boundary.
As for the spinor, it can always be written in
terms of $4d$ Weyl spinors of opposite chirality
\cite{Shuster:1999zf}
\be
\lambda=\begin{pmatrix}\chi\\ \bar{\xi}\end{pmatrix}~.
\ee

Choosing the sign of the mass $\vert
m_{\lambda}\vert=1/2$ we are fixing the chirality of the leading mode
in the near boundary expansion. As we will review, choosing $m_{\lambda}= -
1/2$ the leading mode at the boundary has negative chirality
$\lambda\simeq z^{3/2} \bar{\xi}_{0}$. Using the holographic dictionary,  the GGM functions can then be determined from the renormalized boundary action as
\begin{align}
&C_{0}(k^2)=-\frac{\delta^2 S_{\text{ren}}}{\delta d_{0}\delta d_{0}}\ ,\ \ \ C_{1/2}(k^{2})=-\frac{\bar{\sigma}^{\dot\alpha\alpha}_{i}k^{i}}{2k^2}\frac{\delta^2 S_{\text{ren}}}{\delta \xi_{0}^{\alpha}\delta \bar{\xi}_{0}^{\dot\alpha}} \ ,\ \ \ C_{1}(k^2)=\frac{\eta_{ij}}{3k^2}\frac{\delta^2 S_{\text{ren}}}{\delta a_{0i}\delta a_{j0}}\ , \label{HolographicC}\\
&B_{1/2}(k^2)=- \frac{\epsilon^{\alpha\beta}}{2}\frac{\delta^2 S_{\text{ren}}}{\delta \xi_0^\alpha\delta \xi_0^\beta}\ .\label{HolographicB}
\end{align}
In the remainder of this section, we will derive the results above.

\subsection{Two-point functions from the renormalized boundary action}

In what follows we want to show in some detail how, given a certain background, one can use the procedure of holographic renormalization to calculate two-point functions of current operators (\ref{HolographicC})-(\ref{HolographicB}). The starting point is given by the interactions of the vector multiplet with the background, encoded in the part of the Lagrangian which is quadratic in the vector multiplet.

Near the boundary of the background geometry we assume that all
scalars that might have a non-trivial profile vanish sufficiently fast, so that all their interactions with the
vector multiplet can be neglected in this limit. Therefore, near the boundary the (Euclidean) quadratic Lagrangian approaches that of a vector multiplet minimally coupled to the background
\begin{equation}
L_{\text{quad}}  \underset{\overset{\mathrm{z\to0}}{}}{\simeq} L_{\text{min}}= \frac{1}{2}(G^{\mu\nu}\partial_{\mu}D\partial_{\nu}D-4D^2)+\frac{1}{4}F_{\mu\nu}F^{\mu\nu}+\frac{1}{2}(\bar{\lambda}\slashed{D}\lambda + c.c.) - \frac{1}{2}\bar{\lambda}\lambda \ .\label{quadratic}
\end{equation}
Consequently the equations near the boundary approach
the AdS form
\begin{align}
& (\Box_{AdS}+4)D \underset{\overset{\mathrm{z\to0}}{}}{\simeq} 0\  , \label{AdSscalareom} \\
&(\text{Max})_{AdS}A_{i}  \underset{\overset{\mathrm{z\to0}}{}}{\simeq} 0 \  , \label{AdSvectoreom} \\
&(\slashed{D}_{AdS}- \frac{1}{2})\lambda \underset{\overset{\mathrm{z\to0}}{}}{\simeq} 0 \ .
\end{align}
where we fixed the $5d$ Coulomb gauge in which $A_{z}=0$ and the $4d$ Lorentz gauge $\partial^{i}A_{i}=0$\footnote{Notice that with this gauge choice we can
compute only the coefficient of $\eta_{ij}$ in the vector current
correlator; the additional term can obviously be inferred by
current conservation.}, and the differential operators above take the form
\begin{align}
&\Box_{AdS}=z^{2}\partial_{z}^{2}-3z\partial_{z}+z^{2}(\partial_i)^2  \ , \label{Boxz} \\
&(\text{Max})_{AdS}=z^{2}\partial_{z}^{2}-z\partial_{z}+z^{2}(\partial_i)^2  \ , \label{Maxz} \\
&\ \slashed{D}_{AdS}=z\gamma^{z}\partial_{z}-2\gamma_{z}+z\gamma^{i}\partial_{i}  \label{slashz} \ .
\end{align}
The spinor equation can be rewritten in terms of Weyl components as
\begin{equation}
\begin{cases}
z\partial_{z}\chi+iz\sigma^{i}\partial_{i}\bar{\xi}- \frac{5}{2}\chi \underset{\overset{\mathrm{z\to0}}{}}{\simeq} 0\\
-z\partial_{z}\bar{\xi}+iz\bar{\sigma}^{i}\partial_{i}\chi+\frac{3}{2}\bar{\xi} \underset{\overset{\mathrm{z\to0}}{}}{\simeq} 0  \ .
\end{cases}\label{AdSspinoreom}
\end{equation}
Correspondingly, the asymptotic behavior of the supergravity fields near the boundary takes the following form\footnote{In order to define the logarithmic mode in the bulk we have to introduce an energy scale $\Lambda$ which gets identified with the RG scale of the boundary theory.}
\begin{align}
& D(z,k) \underset{\overset{\mathrm{z\to0}}{}}{\simeq}z^{2}\left( d_0(k)\ln(z\Lambda)+\tilde{d}_0 + O(z^2) \right)  \ , \label{scalingansatzS}\\
& A_{i}(z,k)  \underset{\overset{\mathrm{z\to0}}{}}{\simeq} a_{i0}(k) + z^2(\tilde{a}_{i2}(k) + a_{i2}(k)\ln(z\Lambda)) + O(z^4)  \ , \label{scalingansatzV}\\
&\begin{cases}
\bar{\xi}(z,k)   \underset{\overset{\mathrm{z\to0}}{}}{\simeq}z^{3/2} \left( \bar{\xi}_0(k) + z^2(\bar{\tilde{\xi}}_2(k) + \bar{\xi}_2(k) \ln (z\Lambda)) + O(z^4)\right) \ , \\
\chi(z,k)    \underset{\overset{\mathrm{z\to0}}{}}{\simeq} z^{5/2}\left(\tilde{\chi}_1(k) + \chi_1(k) \ln (z \Lambda)  + O(z^2)\right) \ . \label{scalingansatzSP}
\end{cases}
\end{align}
Note that in the scalar case the leading term at the boundary has a logarithmic scaling. This is a peculiar feature related to the fact that $m^{2}_D=-4$ saturates the stability bound for a scalar field in $\text{AdS}_{5}$ \cite{Breitenlohner:1982bm, Breitenlohner:1982jf}. In the fermionic case the choice $m_\lambda=-1/2$ implies that the leading mode at the boundary has negative chirality, as anticipated.

Every coefficient in the expansion is a function of the momentum $k$, and the variational principle of the supergravity theory is defined by fixing the leading modes at the boundary $(d_0,\, \bar{\xi}_0,\, a_{i0})$ and letting all other coefficients free to vary independently. Substituting the ansatz in the asymptotic equations of motion (\ref{Boxz})-(\ref{slashz}) one can determine the on-shell values of all coefficients but tilded ones (i.e.~the source terms), as local (i.e. polynomial in $k$) functions of the leading modes $(d_0,\, \bar{\xi}_0,\, a_{i0})$. In particular we find
\begin{equation}
a_{i2}=\frac{k^{2}}{2}a_{i0}(k)\ ,\qquad \bar{\xi}_{2}=\frac{k^{2}}{2}\bar{\xi}_{0}(k)\ , \qquad \chi_{1}=-\sigma^{i}k_{i}\bar{\xi}_{0}(k) \ .
\end{equation}
Conversely, the subleading modes in the near boundary expansion are not determined by the near boundary analysis and in general will be non-local functions of the external momenta which can be derived from the exact solutions of the equations of motion.

When evaluated on-shell, the supergravity action \eqref{quadratic} reduces to the boundary terms, which are in general divergent in the limit $z\to0$, and have to be regularized. This can be done considering a cutoff surface $z=\epsilon$, for which,  after Fourier transformation on the $4d$ coordinates, the boundary terms become
\begin{equation}\label{AdSregularized}
S_{\text{reg}}=-\int_{z=\epsilon}\frac{d^{4}k}{(2\pi)^4}\frac{1}{2}\left[\epsilon^{-4}(Dz\partial_zD)_{z=\epsilon}+\epsilon^{-2}(A_{i}z\partial_zA^{i})_{z=\epsilon}-\epsilon^{-4}(\xi\chi+\bar{\chi}\bar{\xi})_{z=\epsilon}\right]\ .
\end{equation}
Note that the fermionic boundary term \cite{Henningson:1998cd,Henneaux:1998ch} is reminiscent of a Dirac mass term.

Plugging the near boundary expansion and taking into account the on-shell relations between the various coefficients, we can collect both a divergent and a finite contribution in the regularized action \eqref{AdSregularized}
\begin{align*}
&S_{\text{reg}}\vert_{\text{div}}= - \int\limits_{z=\epsilon} \frac{d^{4}k}{(2\pi)^4}\frac{1}{2}\ln(\epsilon\Lambda)[2 \ln(\epsilon\Lambda)d_{0}^2+4d_{0}\tilde{d}_0+d_{0}^2+a_{0}^{i}k^{2}a_{0i}+2\xi_{0}\sigma^{i}k_{i}\bar{\xi}_{0}]\ ,\\
&S_{\text{reg}}\vert_{\text{finite}}= - \int \frac{d^{4}k}{(2\pi)^4}\frac{1}{2}[2\tilde{d}^{2}_0+d_{0}\tilde{d}_0+2a_{0}^{i}\tilde{a}_{i2}+a^{i}_{0}\frac{k^2}{2}a_{i0}-(\xi_{0}\tilde{\chi}_{1}+\bar{\tilde{\chi}}_{1}\bar{\xi}_{0})]\ .
\end{align*}
The divergent terms can be subtracted by means of the following covariant counterterms
\begin{equation}
S_{\text{ct}}= - \frac{1}{2}\int\limits_{z=\epsilon}\frac{d^{4}k}{(2\pi)^4}\sqrt{\gamma}[2D^2+\frac{D^2}{\ln(\epsilon\Lambda)}-\frac{1}{2}\ln(\epsilon\Lambda)F_{ij}F^{ij}+2\ln(\epsilon\Lambda)\bar{\lambda}\gamma^{i}k_{i}\lambda]\ .
\end{equation}
The counterterms for the scalar components contribute also to the finite part of the boundary action so we finally get the result
\begin{equation}\label{renormalizedAdS}
S_{\text{ren}}=\frac{N^2}{8\pi^2}\int\frac{d^{4}k}{(2\pi)^4}[d_{0}\tilde{d}_0-2a_{0}^{i}\tilde{a}_{i2}-\frac{1}{2}a^{i}_{0}k^{2}a_{i0}+\xi_{0}\tilde{\chi}_{1}+\bar{\tilde{\chi}}_{1}\bar{\xi}_{0}]\ ,
\end{equation}
where we restored the normalization of the action which was neglected so far. The coefficient is due to the identification $\frac{1}{8\pi G_5} = \frac{N^2}{4\pi^2}$ which we will explain when treating the pure AdS example.

Twice differentiating with respect to the sources we finally get for the two-point functions
\begin{align}
&\langle J(k)J(-k)\rangle=\frac{N^2}{8\pi^2}\left(-  2\frac{\delta \tilde{d}_0}{\delta d_0} \right)                \  , \label{C0} \\
&\langle j_{i}(k)j_{j}(-k)\rangle=\frac{N^2}{8\pi^2} \left(2\frac{\delta \tilde{a}_{i2}}{\delta{a_0^j}} + 2\frac{\delta \tilde{a}_{j2}}{\delta{a_0^i}} +   k^2 \eta_{ij} \right)      \  , \label{C1} \\
&\langle j_{\alpha}(k)\bar{j}_{\dot{\alpha}}(-k)\rangle=\frac{N^2}{8\pi^2}\left( \frac{\delta\tilde{\chi}_{1\alpha}}{\delta\bar{\xi}^{\dot{\alpha}}_0} +  \frac{\delta\bar{\tilde{\chi}}_{1\dot\alpha}}{\delta{\xi}^{\alpha}_0}  \right)   \  , \label{Chalf}\\
&\langle j_{\alpha}(k)j_{\beta}(-k)\rangle = \frac{N^2}{8\pi^2}\left( \frac{\delta \tilde{\chi}_{1\alpha}}{\delta \xi_0^\beta} - \frac{\delta\tilde{\chi}_{1\beta}}{\delta \xi_0^\alpha} \right) \label{B}\ . \
\end{align}
An important comment about the fermionic correlators is in order. From the structure of the spinor equation of motion one can notice that the subleading mode $\tilde{\chi}_1$, which is determined by the full bulk equation, will always have a non-trivial dependence on the leading mode of opposite chirality $\bar{\xi}_0$, ensuring a non-zero value for the function $C_{1/2}$. On the other hand, already at this very general stage, we see that the only way to obtain a non-zero $B_{1/2}$ is to have the mode $\tilde{\chi}_1$ to depend also holomorphically on the source $\xi_0$. In the next section we will show that this is closely related to the presence of Majorana-like couplings in the $5d$ action that can arise only in the presence of a non-trivial profile for R-charged scalars. This should be expected, since a non-zero $B_{1/2}$ requires R-symmetry to be broken. We will see under which conditions non-trivial Majorana-like couplings of the bulk fermions can be produced.


\section{Embedding the hidden sector in a gauged supergravity model}

In this section we will introduce our model for the 5d bulk
theory. As already noticed, the 5d
gravity theory, besides the graviton multiplet, must contain at least
one $\mathcal{N} =2$ vector multiplet, which is dual to the multiplet
of the conserved current of the boundary theory. As a necessary
condition for the theory to be a consistent truncation of  $10d$ type
IIB supergravity,
the matter content must include also one hypermultiplet which is
the $\mathcal{N}=2$ multiplet of the $10d$ dilaton, usually called
universal hypermultiplet. In fact, enlarging the matter content to include an
hypermultiplet is also necessary to the aim of finding interesting
backgrounds, as we will see. Therefore, the minimal 5d content one should consider
consists of $\mathcal{N}=2$ supergravity coupled to a vector multiplet and
a hypermultiplet.

In order to make our program concrete we consider a class of gauged
supergravity theories studied in \cite{Ceresole:2001wi} which actually
contains the minimal field content described above.\footnote{This
  class of theories has the virtue that, for some choices of the
  gauging, the resulting theory is believed to be a consistent
  truncation of the maximally gauged $\mathcal{N} = 8$ supergravity in
  $5d$ (and therefore of $10d$ type IIB compactified on a
  sphere).} We now briefly outline the
main ingredients that specify our Lagrangian, whose form is dictated
by the scalar manifold and the gauging. For further details we refer to
\cite{Ceresole:2001wi,Ceresole:2000jd}. The scalars describe a
non-linear sigma model with target space
\begin{equation}\label{manifold}
 \mathcal{M}=O(1,1)\times \frac{SU(2,1)}{U(2)}.
\end{equation}
The manifold is a direct product of a very special manifold $\mathcal{S} = O(1,1)$ and a quaternionic manifold $\mathcal{Q} = \frac{SU(2,1)}{U(2)}$ spanned by the so-called universal hypermultiplet, which contains the axio-dilaton. The $\mathcal{S}$ factor is parametrized by the vector multiplet real scalar $D$ with metric
\begin{equation}
d s_1^2 = d D^2,
\end{equation}
whereas $\mathcal{Q}$ is parametrized by the four real hyperscalars $q^X = (\phi,\, C_0,\, \eta,\, \alpha)$ with metric
\begin{equation}\label{Qmetric}
d s_2^2 = g_{XY}d q^X d q^Y = \frac{1}{2} \cosh^2(\eta)d \phi ^2 + \frac{1}{2}(2 \sinh^2(\eta) d \alpha + \mathrm{e}^{\phi} \cosh^2(\eta) d C_0)^2 + 2 d\eta^2 \ .
\end{equation}
where $\eta\geq0$ and $\alpha\in[0,2\pi]$. The scalar $\eta$ is sometimes called
squashing mode, since within 10d  compactifications it is related to a squashing parameter of the internal compactification manifold.
The isometries of this
scalar manifold have a $U(2)$ maximal compact subgroup acting on
$\mathcal{Q}$. Since the theory contains two vectors, one in the
gravity multiplet and the other one in the vector multiplet, the
maximal subgroup we can gauge is a $U(1)\times U(1)$. As a minimal set
up we choose to gauge just the $U(1)$ corresponding to the shift
symmetry
\begin{equation}
 \alpha\rightarrow\alpha + c
\end{equation}
of the above metric, which is a compact isometry because the scalar $\alpha$ is  a phase.

The vector field of this $U(1)$ which acts non trivially on the scalar manifold is the graviphoton in the gravity multiplet, so that this gauge symmetry is dual to the R-symmetry of the boundary theory. On the other hand, in our simplified setting the $U(1)$ gauged by the vector belonging to the vector multiplet acts trivially on all supergravity fields (this does not mean that the vector multiplet is free, as we will see when we write its Lagrangian). Notice that the axio-dilaton is neutral under both $U(1)$'s while the complex scalar $\eta e^{i\alpha}$ is charged under the symmetry gauged in the bulk by  the graviphoton. Therefore, a background with a non-trivial profile for the dilaton preserves the R-symmetry, while a non-trivial profile for $\eta$ breaks it. For later reference let us notice that while the axio-dilaton is massless, and holographycally dual to the hidden sector $\mbox{Tr}F_{ij}^2$ operator, the squashing mode $\eta$ has $m^2=-3$ and it is dual to the hidden sector gaugino bilinear. Hence, the leading mode for this field at the boundary would provide an explicit mass to the hidden gauginos  (hence an explicit R-symmetry breaking term), while a subleading term would correspond to a VEV for the gaugino bilinear (hence a spontaneous R-symmetry breaking term).

Starting form our 5d Lagrangian, as already outlined in the previous section, there are basically two  steps one should perform:
\begin{itemize}
\item First, we should  find a non-supersymmetric
  background configuration with just the metric and some of the
  hyperscalars turned on. In order to do this we will truncate the
  Lagrangian to the relevant field content (provided this is
  consistent with the full set of equations) and extract the equations
  of motion which the background must satisfy.
 \item Second, we need to extract the linearized differential
   equations for the vector multiplet fluctuating on the background
   that we will find. To this aim, we will perform a different
   truncation of the Lagrangian setting all fields but the vector
   multiplet to their background values, and retain only the couplings
   which are no more than quadratic in the vector multiplet fields.
\end{itemize}
We will now present the explicit form of these truncated
Lagrangians.

\subsection{Lagrangian for the background}
Let us start by setting to zero the whole vector multiplet, as well as the gravitino, the graviphoton and the fermions of the hypermultiplet. The phase $\alpha$ can be gauge-fixed to zero. The resulting truncated (Euclidean) action reads
\begin{equation}\label{general}
S_{\text{b.g.}} = \int d^{5}x\sqrt{G} \left[ -\frac{1}{2}R + L_{\text{kin}}+ \mathcal{V} \right]	
\end{equation}
where the kinetic term is given in term of the metric \eqref{Qmetric} by $L_{\text{kin}}=\frac{1}{2}g_{XY}\partial_\mu q^X \partial^\mu q^Y$, that is
\begin{equation}
L_{\text{kin}}=\frac{1}{4}\left[4\partial_{\mu}\eta\partial^{\mu}\eta+\cosh^{2}(\eta)\partial_{\mu}\phi\partial^{\mu}\phi+{\rm e}^{2\phi}\cosh^{4}\left(\eta\right)\partial_{\mu}C_{0}\partial^{\mu}C_{0}\right]\ .
\end{equation}
As a consequence of the gauging we have a non-trivial potential given by
\begin{equation}
 \mathcal{V}=\frac{3}{4}\left(\cosh^{2}(2\eta)-4\cosh(2\eta)-5\right).
\end{equation}
We end up with the following system of differential equations
\begin{align}
&R_{\mu\nu}=\frac{2}{3}\mathcal{V}G_{\mu\nu}+2\left(\partial_{\mu}\eta\partial_{\nu}\eta+\frac{1}{4} \cosh^{2}(\eta)\,\partial_{\mu}\phi\partial_{\nu}\phi \right), \label{einstein}\\
&\Box\eta=\frac{1}{2}\frac {\partial \mathcal{V}}{\partial\eta}+\frac{1}{8} \sinh(2\eta)\,\partial_{\mu}\phi\partial^{\mu}\phi ,
\label{chieq}\\
&\Box\phi= - 2 \tanh(\eta)\, \partial_\mu \eta \partial^\mu \phi , \label{dilatoneq}
\end{align}
where $\Box$ is the usual Klein-Gordon operator on a curved space
\begin{equation}
\Box=\frac{1}{\sqrt{G}}\partial_{\mu}(\sqrt{G}G^{\mu\nu}\partial_{\nu})~.
\end{equation}
The condition of asymptotically AdS background can be phrased by taking a metric of the form
\begin{equation}
ds_{5}^{2}= \frac{1}{z^2}\left(dz^2 + F(z) (dx^i)^2 \right)
\label{ansatzgeneric}
\end{equation}
with $F(z)$ approaching 1 at the boundary $z \to 0$. Therefore the solution to the above equations determine the three unknown functions $\phi$, $\eta$ and $F$ of the radial coordinate $z$.

In the case of unbroken R-symmetry, $\eta=0$, the above system of equations reduces exactly to the one considered in \cite{Gubser:1999pk}, and admits both a supersymmetric AdS solution with  constant dilaton, as well as a singular dilaton domain-wall solution  \cite{Gubser:1999pk,Kehagias:1999tr}. The latter breaks both conformal invariance and (all) supersymmetry. Another interesting background is one where also the charged scalar $\eta$ has a non-trivial profile. We will consider all  these examples in turn. Notice that it is only in the R-breaking background   that we need to fix the particular truncation and the form of the gauging which specify our 5d model. This is necessary to derive the potential for $\eta$ and find its profile, and also to find out how it interacts with the fluctuations in the vector multiplet (at least at quadratic level in the vector superfield).

\subsection{Quadratic Lagrangian for the vector multiplet}

We now turn to the action describing the coupling of vector multiplet fluctuations to
the background. To this end we fix $F,\, \phi$ and
$\eta$ to their ($z$-dependent) background value into the full
Lagrangian, and retain only those terms involving the vector
multiplet up to second order.  The resulting (Euclidean) action can be
divided in two pieces
\begin{equation}
S_{\text{quad}} =  \int d^{5}x\sqrt{G}\left[ L_{\text{min}} + L_{\text{int}}\right] . \label{quadratic2}
\end{equation}
The first one contains kinetic terms and mass terms for the
fluctuations, and it is uniquely fixed by the dimensions of the dual
operators and their minimal coupling to the metric to be
$L_{\text{min}}$, eq.~ (\ref{quadratic}). The second one contains interactions with the scalars $\phi$ and
$\eta$ and takes the form
\begin{align}
L_{\text{int}} & =  \frac{1}{2} \delta M^2 D^2  - \delta m_D
\bar{\lambda}\lambda \nonumber \\
& - \frac{1}{2} \left(m_M \bar{\lambda}\lambda^c + v_M \bar{\lambda} (\slashed{\partial}\eta)\lambda^c + \tilde{v}_M \bar{\lambda} (\slashed{\partial}\phi)\lambda^c + c.c. \right).\label{lint}
\end{align}
where
\begin{align}
& \delta M^2 = 2(\cosh^2(2\eta) - \cosh(2\eta)) \ , \ \delta m_D = - \frac 12 \sinh^2(\eta) \label{ceta1}\\
& m_M =  i \sinh(\eta) \ , \ v_M = -\frac{i}{\cosh(\eta)} \ , \ \tilde{v}_M =\frac{i}{2} \sinh(\eta). \label{ceta2}
\end{align}
In the first line there are ($z$-dependent) shifts for scalar mass squared
and Dirac fermion mass, whereas in the
second line there are a Majorana mass term and additional
Majorana-like couplings. We wrote the couplings in a 5d covariant
manner, but one should bear in mind that $\eta$ and $\phi$ are background
values which actually can depend only on the radial coordinate,
so that the additional terms are equivalent to $4d$ covariant terms
constructed with a $\gamma_5$ matrix. Notice that all couplings (\ref{ceta1})-(\ref{ceta2}) vanish if
$\eta$ is identically zero in the background.

From the action \eqref{quadratic2} we get the equations of motions
\begin{align}
& (\Box+4-\delta M^2)D=0 \ , \label{scalarminimal} \\
& \frac{1}{\sqrt{G}}\partial_{\mu}(\sqrt{G}G^{\mu\rho}G^{\nu\sigma}F_{\rho\sigma})=0\ , \label{vectorminimal} \\
&(\slashed{D} - \frac{1}{2} - \delta m_D)\lambda - (m_M + v_M \slashed{\partial}\eta + \tilde{v}_M \slashed{\partial}\phi )\lambda^c = 0 \ , \label{spinorminimal}
\end{align}
where
\begin{align}
& \ \Box = \frac{1}{\sqrt{G}}\partial_{\nu}(\sqrt{G}G^{\mu\nu}\partial_{\mu}) \ , \\
& \ F_{\mu\nu} = \partial_{\mu}A_{\nu}-\partial_{\nu}A_{\mu}  \ , \\
& \ \slashed{D} = e^{\mu}_{a}\gamma^{a}\left(\partial_{\mu}+\frac{1}{8}\omega_{\mu}^{cb}[\gamma_{b},\gamma_{c}]\right).
\end{align}
As already noticed,
the 5d spinor is equal in form to a 4d Dirac spinor and it is
often useful to rewrite its equation of motions in terms of chirality
eigenstates, that is
\be
\lambda=\begin{pmatrix}\chi\\ \bar{\xi}\end{pmatrix}, \qquad
\bar{\lambda}=\Lorenzo{-}\begin{pmatrix}\xi & \bar{\chi}\end{pmatrix}, \qquad \lambda^c=\begin{pmatrix}\xi\\ \Lorenzo{-}\bar{\chi}\end{pmatrix}.\label{chiralityeigen}
\ee
In terms of Weyl components $\chi$ and $\xi$, eq.~(\ref{spinorminimal}) becomes
\begin{align}
& (z\partial_z - \frac 52 + z\frac{F'}{F}-\delta m_D)\chi +i \frac{z}{\sqrt{F}} \sigma^i\partial_i \bar \xi
  - (m_M + v_M z\eta' + \tilde{v}_M z\phi' )\xi=0~,\\
& (z\partial_z - \frac 32 + z\frac{F'}{F}+\delta m_D)\bar \xi -i \frac{z}{\sqrt{F}}\bar \sigma^i\partial_i\chi- (m_M - v_M z\eta' - \tilde{v}_M z\phi' ) \bar \chi=0~.
\end{align}
As can be seen from above equations, when Majorana-like
couplings are turned on, not only $\bar \xi$ but also $\xi$ appears in the equation
for $\chi$, and vice-versa. As we concluded in the previous section,
we thus see that it is necessary to turn on a background for the
scalar $\eta$ in order to have correlators with a non-zero $B_{1/2}$.

\subsection{Renormalized action with a non-trivial $\eta$}

When the scalar $\eta$ has a non-trivial profile and has non-vanishing leading boundary behavior, the renormalized action for the
vector multiplet  should be slightly modified.

The scalar $\eta$ has $m^2=-3$, and therefore its leading and subleading boundary behavior
is
\be
\eta \underset{\overset{{z\to0}}{}}{\simeq} \eta_0 z + \tilde \eta_2
z^3 + \dots \label{etabound}
\ee
As the numerical analysis in the following sections will show, whenever the leading mode $\eta_0$ is present (a source term for the corresponding $\Delta=3$ boundary operator,
the hidden gaugino bilinear), the renormalized boundary action
(\ref{renormalizedAdS}) should be modified by the following term
\begin{equation}
\label{etaren+}
S_{\text{ren}}^\eta=\frac{N^2}{8\pi^2}\int\frac{d^{4}k}{(2\pi)^4}[i\eta_0(\xi_0\xi_0-\bar\xi_0\bar\xi_0)]\ .
\end{equation}
Accordingly, the expression for the correlator (\ref{B}) is
modified to
\begin{equation}
\langle j_{\alpha}(k)j_{\beta}(-k)\rangle_{\eta} =  \frac{N^2}{8\pi^2}\left( \frac{\delta \tilde{\chi}_{1\alpha}}{\delta \xi_{0\beta}} - \frac{\delta\tilde{\chi}_{1\beta}}{\delta \xi_{0\alpha}} +2i\epsilon_{\alpha\beta}\eta_0\right).\label{jjcorrelchimod}
\end{equation}
The
corrected expression (\ref{jjcorrelchimod}) is necessary to ensure
that the fermionic
correlator properly goes to zero at large momenta, as dictated by supersymmetry restoration at high energy. The ultra-local term
(\ref{etaren+}) can be seen as a counterterm which we add to the boundary action in order to reabsorb an unwanted contact term in the correlator.  This countertem only depends on quantities that are
held fixed in the variational principle.

Notice that if the $\eta$ profile has a leading boundary behavior proportional to $\tilde \eta_2$, which is holographically dual to a purely dynamical generation of an R-symmetry
breaking VEV, no modification in the renormalized boundary action
occurs. Still, having $\eta$ a non trivial profile, $\tilde\chi_1$ would
depend on $\xi_0$, and hence the correlator (\ref{B}) would be in general different from zero.

The origin of  the additional term (\ref{etaren+}) can alternatively be motivated as follows. The interaction Lagrangian (\ref{lint}) at linear order in $\eta$ reads
\be
L_\mathrm{int}^\mathrm{lin} =  \frac{1}{2} \left(-i\eta
\bar{\lambda}\lambda^c + i \bar{\lambda}
(\slashed{\partial}\eta)\lambda^c -\frac{i}{2}\eta \bar{\lambda}
(\slashed{\partial}\phi)\lambda^c + c.c. \right)~, \label{lintlin}
\ee
where we can actually neglect the third term, since in
a background with a non-trivial dilaton profile, which necessarily behaves as $z\partial_z \phi
={\cal O}(z^4)$, this cannot contribute to the boundary action.

The key observation is that the following boundary term
\be
S_{\text{reg}}^\eta=\int_{z=\epsilon}\frac{d^{4}k}{(2\pi)^4}\frac{i}{2}
\epsilon^{-4} \left[\eta (\xi\xi
  -\bar\xi\bar\xi-\chi\chi+\bar\chi\bar\chi)\right]_{z=\epsilon}\ .
\ee
is obtained if one integrates by parts the second  term in (\ref{lintlin}). We note that this boundary term is now Majorana-like, in contrast with the usual one, eq.~(\ref{AdSregularized}), which is Dirac-like.
The term bilinear in $\chi$ is always vanishing at the boundary, but
we notice that when $\eta \sim \epsilon$ the term bilinear in $\xi$ is
actually finite, and is exactly the term
(\ref{etaren+}) after we restore the proper normalization.

One can easily verify that the action for $\eta$ and $\lambda$ with the interactions given in (\ref{lintlin}) vanishes on-shell up to quartic terms in those fields.  Therefore, to this order of approximation, supplementing the renormalized boundary action by the counterterm (\ref{etaren+}) is equivalent to considering an interaction Lagrangian modified with respect to (\ref{lintlin}) by replacing the derivative interaction with the one obtained after integration by parts.

\section{Holographic correlators in AdS}

As a warm up exercise we want to compute the GGM two-point functions
for a pure AdS background, which is a solution of eqs.(\ref{einstein})-(\ref{dilatoneq}) with $\phi=\eta=0$. This exercise has several
motivations. First of all it will enable us to verify that our machinery correctly
reproduces what we expect from a conformal and supersymmetric case,
namely eqs.~(\ref{susycandb}). Second, the values for the
correlators that we find in AdS will be the reference  to
confront with, when considering other backgrounds. In particular,
each correlator will have to
asymptote to those of the pure AdS case, at large momenta. Finally, the
computations we perform in this section can be of interest in a different
context, that is when conformality and supersymmetry breaking are
implemented by a hard wall in AdS (for this perspective, see
\cite{hwpaper}).

The pure AdS solution is a trivial solution of our 5d effective model. However, in order to fix the overall normalization of correlators, it is useful
to uplift it to the $AdS_5 \times S^5$
solution of $10d$ type IIB supergravity, which reads (see e.g.~\cite{Gubser:1999pk})
\begin{align}
&ds^{2}_{10}=\frac{L^2}{z^{2}}(dz^{2}+(dx^i)^2)+L^2d\Omega_{5}^{2}\ , \label{pureAdS}\\
&F_{5}=\frac{N\sqrt{\pi}}{2\pi^{3}}(\text{vol}(S_{5})+\frac{1}{z^{5}}d^{4}x\wedge dz)\ ,
\end{align}
where the radius of $AdS_{5}$ is fix to be $L^{4}=\frac{k_{10}N}{2\pi^{5/2}}$ by 10d Einstein equations. The overall constant in front of the $10d$ action is $1/2k_{10}^2$ so that, substituting the value of the $10d$ Newton constant in terms of the string theory parameters $k_{10}=\sqrt{8\pi G_{10}}=8\pi^{7/2}g_{s}\alpha^{\prime2}$, we get $L^{4}=4\pi g_{s}N\alpha^{\prime2}$. Taking $L=\alpha^{\prime}=1$ we find $G_{5}=\frac{\pi}{2N^2}$ and the overall constant in front of the $5d$ effective action is $1/8\pi G_{5}=\frac{N^2}{4\pi^2}$.

In pure AdS the equations of motion
(\ref{AdSscalareom}), (\ref{AdSvectoreom}) and  (\ref{AdSspinoreom}) are related to standard Bessel equations \cite{Bessel} and therefore it is possible to get
analytic solutions for the fields
\begin{align}
&D(z,k)=-z^{2}K_{0}(kz)d_{0}(k) \label{AdSexact1}\ ;\\
&A_{i}(z,k)=zkK_{1}(kz)a_{0i}(k) \label{AdSexact2}\ ;\\
&\bar{\xi}(z,k)=z^{5/2}kK_{1}(kz)\bar{\xi}_{0}(k)\ ,\ \chi=-z^{5/2}\sigma^{i}k_{i}K_{0}(kz)\bar{\xi}_{0}(k)\ .\label{AdSexact3}
\end{align}
The modified Bessel functions $K_{\nu}(x)$ can be written
as power series which contain logarithmic modes for integer
$\nu$ \cite{Bessel}. For our concerns, all we need to know is
the behaviour of these functions near the origin
\begin{align}
&K_{0}(x)\underset{\overset{{x\to0}}{}}{\simeq} - \ln x+\ln 2-\gamma+O(x^2)\ , \label{Besselboundary1}\\
&K_{1}(x)\underset{\overset{{x\to0}}{}}{\simeq} \frac{1}{x}\left[1+\frac{x^2}{4}(2\ln x-2\ln 2+2\gamma-1)+O(x^4)\right] \label{Besselboundary2}\ .
\end{align}
 Using these expansions we get
\begin{align}
&\tilde{d}_{0}(k)=\left[-\frac{1}{2}\ln\left(\frac{\Lambda^2}{k^2}\right)-\ln2+\gamma\right]d_{0}(k)\ , \label{subAdS1}\\
&\tilde{a}_{i2}(k)=\frac{k^2}{2}\left[-\frac{1}{2}\ln\left(\frac{\Lambda^2}{k^2}\right)-\ln2+\gamma-\frac{1}{2} \right]a_{i0}(k)\ ,\label{subAdS2}\\
&\bar{\tilde{\xi}}_{2}(k)=\frac{k^2}{2}\left[-\frac{1}{2}\ln\left(\frac{\Lambda^2}{k^2}\right)-\ln2+\gamma-\frac{1}{2} \right]\bar{\xi}_{0}(k)\ , \label{subAdS3}\\
&\tilde{\chi}_{1}(k)=\left[-\frac{1}{2}\ln\left(\frac{\Lambda^2}{k^2}\right)-\ln2+\gamma\right]\sigma^{i}k_{i}\bar{\xi}_{0}(k)\ .\label{subAdS4}
\end{align}
Substituting these expressions into eqs.(\ref{C0})-(\ref{B}), we get for the two-point functions
\begin{align}
&\langle J(k)J(-k)\rangle=\frac{N^2}{4\pi^2}\left[\frac{1}{2}\ln\left(\frac{\Lambda^2}{k^2}\right)+\ln2-\gamma\right]\ ; \label{AdSC}\\
&\langle j_{i}(k)j_{j}(-k)\rangle=-\frac{N^2}{4\pi^2}\left(\eta_{ij}-\frac{k_{i}k_{j}}{k^2}\right)k^2\left[\frac{1}{2}\ln\left(\frac{\Lambda^2}{k^2}\right)+\ln2-\gamma\right]\ ;\\
&\langle j_{\alpha}(k)\bar{j}_{\dot{\alpha}}(-k)\rangle=\frac{N^2}{4\pi^2}\sigma^{i}k_{i}\left[\frac{1}{2}\ln\left(\frac{\Lambda^2}{k^2}\right)+\ln2-\gamma\right]\\
&\langle j_{\alpha}(k)j_{\beta}(-k)\rangle=0\ . \label{AdSB}
\end{align}
Our results are in agreement with CFT computations \cite{Muck:1998rr, Mueck:1998iz}. Note that we can always subtract the constant contribution $\ln2-\gamma$ to the two-point functions by means of finite counterterms which preserve the $\mathcal{N} =2$ supersymmetry of the bulk action, so these terms are inessential and will be ignored in what follows.

As expected for a supersymmetric background we find that the relations
(\ref{susycandb}) are satisfied, and thus that both gaugino
\eqref{gauginomass} and sfermion
masses \eqref{sfermionmass} are identically zero. In a general
superconformal theory the OPE of the conserved current satisfies some
general constraints which were studied in general in
\cite{Fortin:2011nq} and applied to the GGM formalism in
\cite{Fortin:2011ad}. In particular, if the hidden sector is exactly
superconformal as it is the case for $\mathcal{N} =4$ SYM, only the unit
operator in the OPE of $J(x)J(0)$ can have an expectation value,
leading to
\begin{equation}
C_{0}(x)=C_{1/2}(x)=C_{1}(x)=\frac{\tau}{16\pi^{4}x^4}\rightarrow C_{0}(k^2)=C_{1/2}(k^2)=C_{1}(k^2)=\frac{\tau}{16\pi^2}\ln\left(\frac{\Lambda^2}{k^2}\right)\ \label{finalresult} ,
\end{equation}
where the coefficient $\tau$ associated to the unit operator has been
exactly determined from 't~Hooft anomaly in \cite{Anselmi:1997ys} and
gives the contribution of the CFT matter to the beta function
associated to the gauge coupling constant of the $U(1)$ subgroup of
$SO(6)$ that we are gauging. For the $AdS_{5}$ case we find $\tau=2
N^2$. We note here that such a large number would be in contrast with
keeping the SSM gauge couplings perturbative before unification. We
will not comment on this further, besides saying that we are really
trying to extract from this holographic approach qualitative features
of correlators in strongly coupled hidden sectors, that we assume are
a good approximation even outside the large $N$ limit.

\section{Holographic correlators in a dilaton-domain wall}

In this section we do a step further and apply our machinery to a supersymmetry breaking background, which is also a solution of our 5d supergravity
Lagrangian. In this case we keep a trivial profile for the squashing mode, $\eta=0$, but allow for a non-trivial dilaton profile. We will see how the IR behavior of the correlators will change drastically with respect to their conformal expressions found in the previous section.

The dilaton-domain wall is in fact a solution of the full 10d type IIB
supergravity found in \cite{Gubser:1999pk, Kehagias:1999tr}. This is a
singular solution with a non-trivial background for the dilaton $\phi$
which preserves the full $SO(6)$ R-symmetry. Upon dimensional reduction
on $S^5$ we get the following $5d$ background
\begin{align}
&ds_{5}^{2}=ds_{5}^{2}= \frac{1}{z^2}(dz^2+{\sqrt{1-z^{8}}}\,(dx^i)^2)\label{Gubser3}\ ,\\
&\phi (z)= \phi_\infty + \sqrt{6}\, \mathrm{arctanh} (z^4)\ .
\end{align}
The metric goes to $AdS_{5}$ at the boundary $z\to 0$ and presents a naked singularity in the deep interior of the bulk, which we have set to $z=1$ by adjusting one of the constants of integration. At the singularity the dilaton diverges
\begin{equation}
\lim_{z\to1}\phi(z) = \infty\ .
\end{equation}
The presence of the naked singularity signals a breakdown of the supergravity approximation and therefore the holographic interpretation of this background as a well-defined field theory could be problematic. It appears that this particular singularity is physically acceptable according to the two criteria  of  \cite{Gubser:2000nd} and \cite{Maldacena:2000mw}. Respectively, its scalar potential is bounded from above (it is exactly zero), and $g_{tt}$ is monotonously decreasing towards the singularity. The reason this solution has had some bad reputation is due to the fact that it fails another criterium put forward in \cite{Gubser:2000nd}, namely that it has no generalization with a horizon.

A possible physical interpretation of this background was discussed in
\cite{Gubser:1999pk, Constable:1999ch}.
Suffices here to say that it describes a vacuum of a theory which in
the UV coincides with ${\cal N}=4$ SYM,
where however a non-trivial VEV for $\tr F_{ij}^2$ is turned on triggering
confinement and SUSY breaking.
In the following we will probe some of its features by the explicit computation of the GGM correlators. This background is interesting for our program because it breaks, besides conformality,  all the supersymmetries (as one can see from the supersymmetry transformation of the dilatino) and it preserves the $SO(6)$ symmetry, so that we can consider an $\mathcal{N} = 2$ vector multiplet gauging a $U(1)\subset SO(6)$.

The effective action at the linearized level for the $\mathcal{N} =2$ vector multiplet in the dilaton-domain wall is of the form \eqref{quadratic}, and the resulting equations of motion will take the schematic form
\begin{align}
&(\Box_{DW}-4)D\equiv\left(z^{2}\partial_{z}^{2}-\left(\frac{3+5z^{8}}{1-z^{8}}\right)z\partial_{z}+\frac{z^{2}(\partial_i)^2}{\sqrt{1-z^{8}}}-4\right)D=0\ ,\label{ddwS}\\
&(\text{Max})_{DW}A_i\equiv \left(z^{2}\partial_{z}^{2}-\left(\frac{1+3z^{8}}{1-z^{8}}\right)z\partial_{z}+\frac{z^{2}(\partial_i)^2}{\sqrt{1-z^{8}}}\right)A_i=0\ ,\label{ddwV}\\
&(\slashed{D}_{DW}-\frac12)\lambda\equiv\left(z\gamma_z\partial_{z}-2\frac{1+z^{8}}{1-z^{8}}\gamma_z+\frac{z}{(1-z^{8})^{1/4}}\gamma^{i}\partial_{i}-\frac12\right)\lambda=0\ .\label{ddwSP}
\end{align}
We note that the AdS equations are modified by terms of $O(z^8)$ in a near boundary expansion.

The second order equations for the fluctuations of the supergravity fields can be solved once two boundary conditions are specified.\footnote{For the sake of the argument that follows, we can convert the two first order equations for the spinors $\chi$ and $\bar\xi$ into a single second order equation for $\bar\xi$.}  One boundary condition will always determine the leading term at the boundary, fixing the overall normalization of the solution. The second condition should be a regularity condition in the bulk. In the case under consideration this means to fix the behavior near the singular point $z = 1$.

Expanding eqs.~ (\ref{ddwS})--(\ref{ddwSP}) to the leading order in $1-z\equiv y \to 0$ we get
\begin{align}
& (y^2\partial_y^2  + y \partial_y) D = 0 \ , \\
& (y^2\partial_y^2 + \frac{1}{2} y \partial_y )A_i = 0 \ , \\
& (y^2\partial_y^2  + \frac{5}{4} y \partial_y  - \frac{1}{8} )\bar{\xi} = 0 \ .
\end{align}
whose solutions are given in terms of two undetermined coefficients $\alpha$ and $\beta$ as
\begin{align}
& D  \underset{\overset{{y\to0}}{}}{\simeq}  \alpha_0 \ln y + \beta_0  \ , \\
& A_i \underset{\overset{{y\to0}}{}}{\simeq} \alpha_{i1} + \beta_{i1} y^{1/2}  \ , \\
& \bar{\xi} \underset{\overset{{y\to0}}{}}{\simeq} \alpha_{1/2} y^{-1/2} + \beta_{1/2} y^{1/4} \ .
\end{align}
The differential equations are well posed if we require, for all of the three fields, that a linear combination of $\alpha$ and $\beta$ vanishes.\footnote{For instance $D=0$ or $\partial\bar{\xi} = \text{const.}$ at the singularity  are not suitable boundary conditions because they would kill both the coefficients.} A condition giving a unequivocal choice for all of the three fields is requiring that both the field and its derivative are finite at the singularity. This condition can be satisfied for all of the three fields and their first derivatives, except for the first derivative of the fermion, which will diverge in any case. We thus select the choice of parameters $\alpha_0 = \beta_1 = \alpha_{1/2} = 0$.\footnote{More general choices of the boundary conditions are in principle allowed (in bottom-up approaches for instance), and would give rise to different physics. For this perspective see \cite{hwpaper}.}

Once we specify the boundary conditions, a solution to eqs.~(\ref{ddwS})--(\ref{ddwSP}) can be found numerically for any value of the parameter $k$ corresponding to the $4d$ momentum.
By using the holographic formulas (\ref{C0})--(\ref{Chalf}) we can then plot the $C_s$ functions.

\begin{figure}
\begin{center}
\includegraphics[height=0.20\textheight]{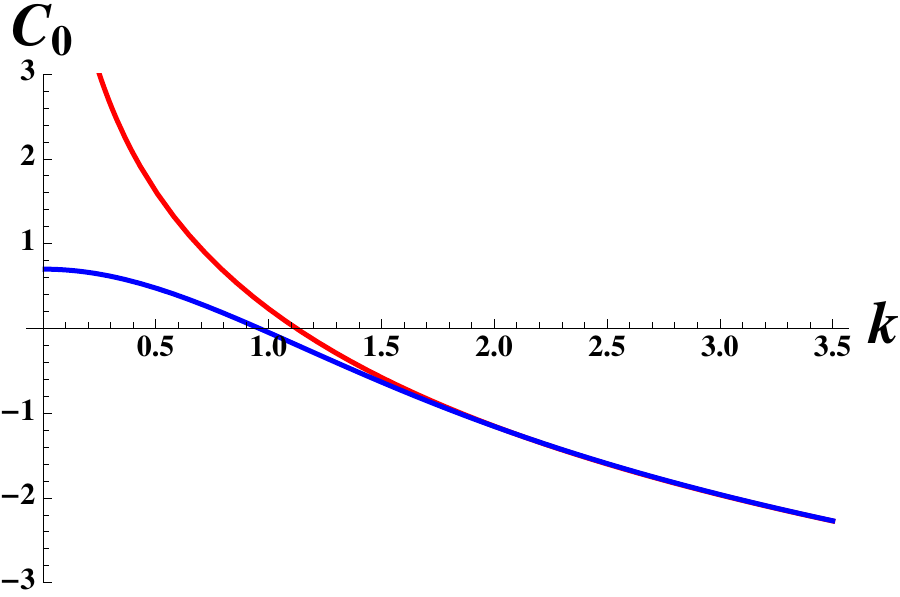}
\caption{\small $C_0$ function: in red the AdS logarithm, in blue the dilaton domain wall result.\label{plotczero}}
\end{center}
\end{figure}

\begin{figure}
\begin{center}
\includegraphics[height=0.20\textheight]{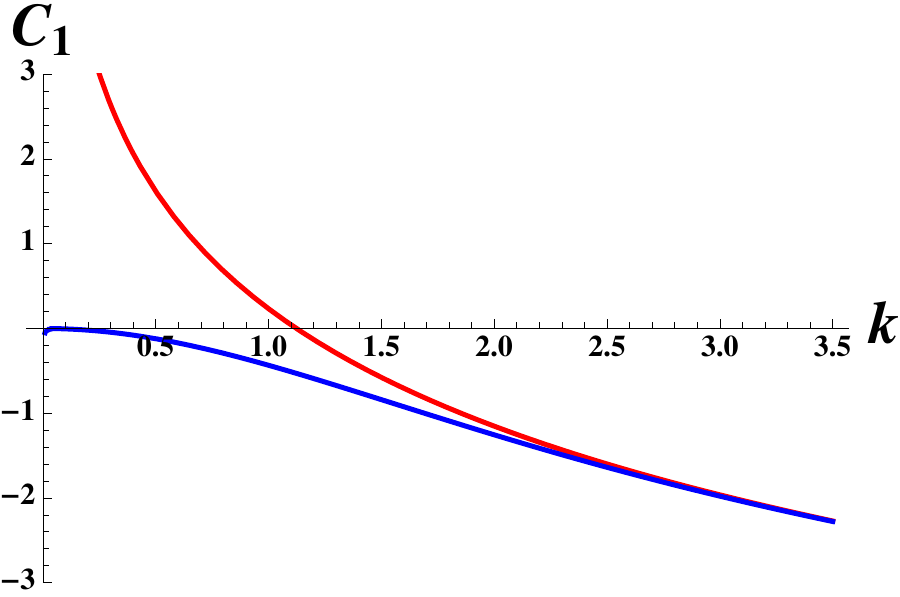}
\caption{\small $C_1$ function: in red the AdS logarithm, in blue the dilaton domain wall result.\label{plotcone}}
\end{center}
\end{figure}

\begin{figure}
\begin{center}
\includegraphics[height=0.20\textheight]{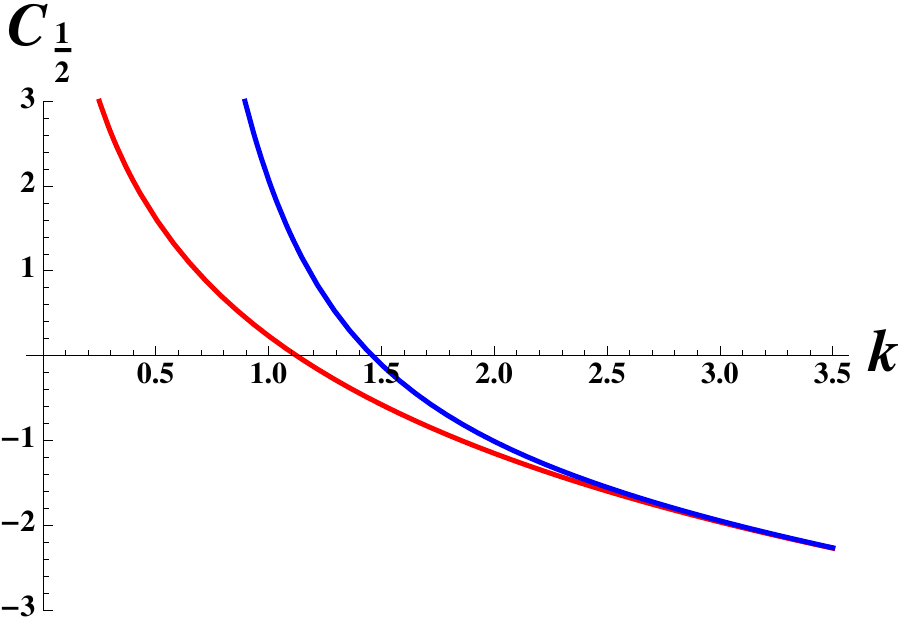}
\caption{\small $C_{1/2}$ function: in red the AdS logarithm, in blue the dilaton domain wall result.\label{plotchalf}}
\end{center}
\end{figure}

\begin{figure}
\begin{center}
\includegraphics[height=0.20\textheight]{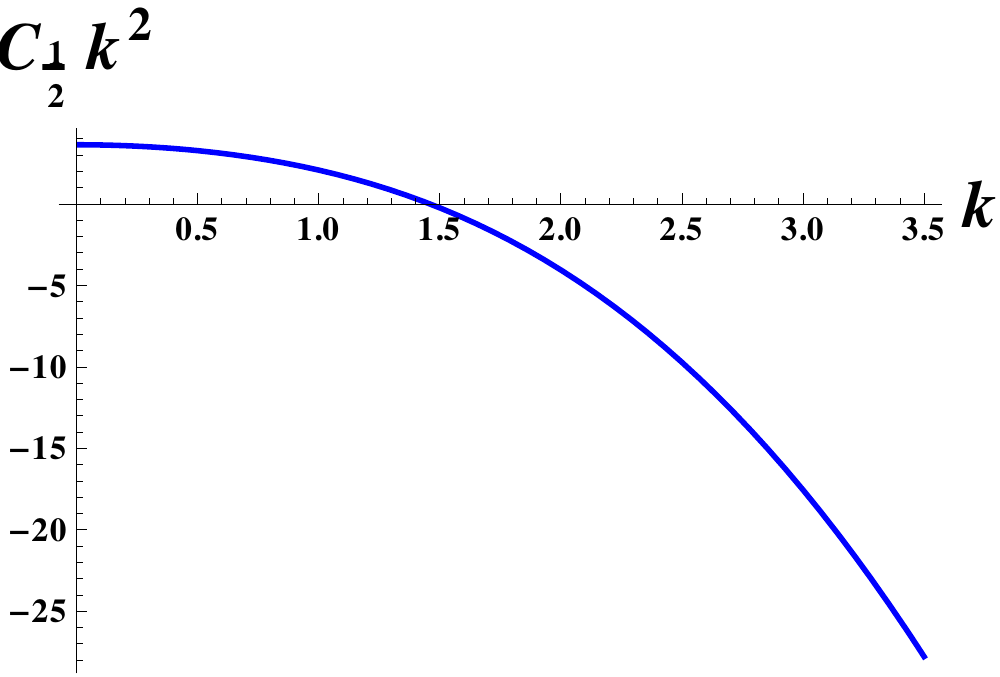}
\caption{\small $k^2C_{1/2}$: for $k\to 0$ it goes to a finite value, indicating that there is a $1/k^2$ pole in $C_{1/2}$ for the dilaton domain wall.\label{plotpole}}
\end{center}
\end{figure}

We show the plots in figures \ref{plotczero}, \ref{plotcone} and \ref{plotchalf}. In each graph we plot both the result for the supersymmetric AdS case, as well that for the dilaton domain wall solution. It is reassuring to note that the AdS tails for the three graphs correctly coincide and tend to their supersymmetric value.

One of the interesting results of the plots is the $k^{-2}$ IR behavior of the fermionic correlator $C_{1/2}$. In figure \ref{plotpole}  we plot $k^2C_{1/2}$, which clearly shows this correlator has a $1/k^2$ pole at zero momentum. This kind of behavior is related to the existence of massless excitations carrying the same quantum numbers of the corresponding current. For the fermionic current $j_\alpha$, this signals the existence of massless fermions, tipically 't Hooft fermions, that compensate the global anomaly of the unbroken $U(1)_R$-symmetry \cite{Buican:2009vv}. Note, in passing, that imposing the ``wrong" boundary condition for the vector field fluctuations, namely $\alpha_1=0$, we would have gotten a $k^{-2}$ pole also for $C_1$. For the vector current
this would be a massless Goldstone boson.
This would imply the existence of Goldstone bosons associated to some broken global symmetry, which cannot be the case here since the original 10d background preserves the full
$SO(6)$ (and hence our $U(1)\times U(1)_R$) symmetry.

While we cannot prove that there are indeed R-charged 't Hooft fermions in our strongly coupled theory, and just observe that the holographic analysis suggests them to be there, it is useful to refer to the full 10d background to get some more confidence about our result. From the 10d perspective there is a whole $SO(6)$ symmetry which the background preserves. Hence, at every scale there must exist
massless fermions in the spectrum so to match the UV global
anomaly. The UV fixed point is $\mathcal{N} =4$ SYM, which has indeed a
non-zero global anomaly for the $SO(6)$ current. At this point one may
think that the $U(1)$ global symmetry of the hidden sector which we
are eventually weakly gauging and identifying with the (simplified) SM gauge
group is anomalous. This has not to be the case because having a non-zero
$SO(6)^3$ anomaly still allows to consider a non-anomalous $U(1)$
subgroup inside $SO(6)$. On the contrary, our result suggests that (part of) the $SU(4)$ anomaly is transmitted to the
$U(1)_R$  current. Let us emphasize that any other anomalous global symmetry would not provide a pole to the fermionic correlator $C_{1/2}$, which is neutral under any global symmetry but the R-symmetry. Hence, field theory expectations would suggest that when the R-symmetry is broken, R-charged 't Hooft fermions would not exist, and the pole in the fermionic correlator should vanish. We will come back to this point in the next section.

The Majorana gaugino mass, determined by
$B_{1/2}$ through (\ref{gauginomass}), vanishes because of unbroken
R-symmetry. However, the pole in $C_{1/2}$ provides for a Dirac mass
for the SSM gaugino. This is very similar to any other model of
R-symmetric Dirac gaugino masses, except that the massless fermion in
the adjoint that must couple bilinearly with the gaugino is here a
composite fermion generated at strong coupling. The
phenomenological aspects of this spectrum will be discussed in more
detail in \cite{hwpaper}. Suffices to say, here, that the soft spectrum, in this situation, is very much reminiscent of that of gaugino mediation models. See \cite{Benakli:2008pg,Intriligator:2010be} for a discussion of Dirac gaugino masses in General Gauge Mediation.

Let us finally  notice how different are the $C_s$ in the dilaton domain wall
background with respect to the ones in AdS, at large
momentum. Numerically we find that
\be
C_0-4C_{1/2}+3C_1 \sim O(k^{-8}), \quad k\to \infty\ .\label{asymptc}
\ee
This is due to the fact that the correction of the domain
wall metric
with respect to the AdS one near the boundary is of $O(z^8)$. Note
that since the dilaton does not enter the equations for the vector
multiplet fluctuation, its $O(z^4)$ behaviour near the boundary does
not influence the $C_s$. Another nice feature of the asymptotic
behaviour (\ref{asymptc}) is that it makes the integral (\ref{sfermionmass})
nicely convergent in the UV.

Both the IR and the UV limit of the $C_s$ functions could also be
determined analytically by studying the
equations (\ref{ddwS})--(\ref{ddwSP}) in the respective limits $k\to
0$ and $k\to \infty$. We delay this study to \cite{hwpaper}.

\section{Holographic correlators in a dilaton/$\eta$-domain wall}

Let us discuss our last example, and look for a solution of eqs.(\ref{einstein})-(\ref{dilatoneq}) with a non-trivial profile for both the dilaton
and the squashing mode. The latter breaks the R-symmetry so one should expect a very different behavior for the correlators.

In fact, in what follows  we will only turn on a perturbative profile for the R-symmetry breaking
scalar $\eta$, that is we
consider only the linearized equation for $\eta$ on the dilaton domain
wall background, and neglect the backreaction of such a
profile on the dilaton and the metric. As we are going to show, this will still be enough to provide a drastic change
in the holographic correlators (nicely matching, again, field theory expectations).

The linearized equation for $\eta$ is most conveniently written and solved using the following parametrization of the asymptotic AdS metric (with boundary at $r \to \infty$)
\begin{equation}
 ds^2 =\left( dr^2 +  e^{2r}(dx^i)^2 \right)
\end{equation}
and reads
\begin{equation}
\eta''(r) + 4 \mathrm{coth}(4r)\eta'(r)+3\eta(r)-\dfrac{3}{2(\mathrm{sinh}(4r))^2}\eta(r)=0.
\end{equation}
The solution depends on two integration constants $A$ and $B$ and is given by
\begin{align}
\eta(r) =( e^{8r}-1)^{\frac{1}{4}\sqrt{\frac{3}{2}}}&\left[ A {}_{\phantom{1}2}F_1\left(\frac{2+\sqrt{6}}{8},\frac{4+\sqrt{6}}{8},\frac{3}{4},e^{8r}\right)  \nonumber \right.\\
& \left.+  B {}_{\phantom{1}2}F_1\left(\frac{4+\sqrt{6}}{8},\frac{6+\sqrt{6}}{8},\frac{5}{4},e^{8r}\right)\right]\ ,
\end{align}
where ${}_{\phantom{1}2}F_1$ is the hypergeometric function.

Changing variables to the usual $z=e^{-r}$ radial coordinate, one can verify that indeed this solution has the expected  behavior (\ref{etabound}) near the boundary, with $\eta_0$ and $\tilde{\eta}_2$ expressed as linear combinations of $A$ and $B$. On the other hand, studying the equation near the singularity $y=1-z \to 0$ one finds the following behavior
\be
\eta \underset{\overset{{y\to0}}{}}{\simeq} \alpha y^{\frac{1}{4}\sqrt{\frac{3}{2}}}+\beta y^{-\frac{1}{4}\sqrt{\frac{3}{2}}}\ ,
\ee
with $\alpha$ and $\beta$ which are in turn linear combinations of $A$ and $B$. If one imposes the boundary condition at the singularity so to meet the criterion on the boundedness of the potential \cite{Gubser:2000nd}, that is $\beta=0$,  one finds a relation between $A$ and $B$ which imposes both $\eta_0$ and $\tilde{\eta}_2$ to be turned on at the boundary (indicating that R-symmetry is broken explicitly in the hidden sector). This implies that in doing the holographic renormalization procedure one should bear in mind the discussion in section 3.3 and augment the boundary action by the term (\ref{etaren+}).

\begin{figure}
\begin{center}
\includegraphics[height=0.20\textheight]{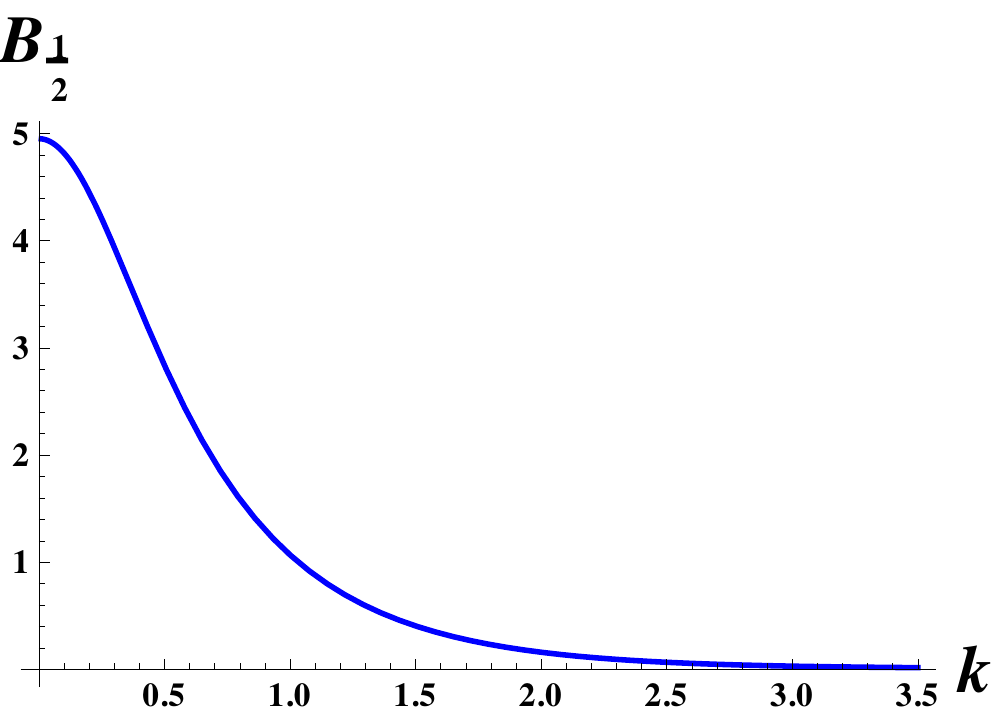}
\caption{\small $B_{1/2}$ obtained on a background with a non-trivial profile for $\eta$. \label{BHalf}}
\end{center}
\end{figure}

\begin{figure}
\begin{center}
\includegraphics[height=0.20\textheight]{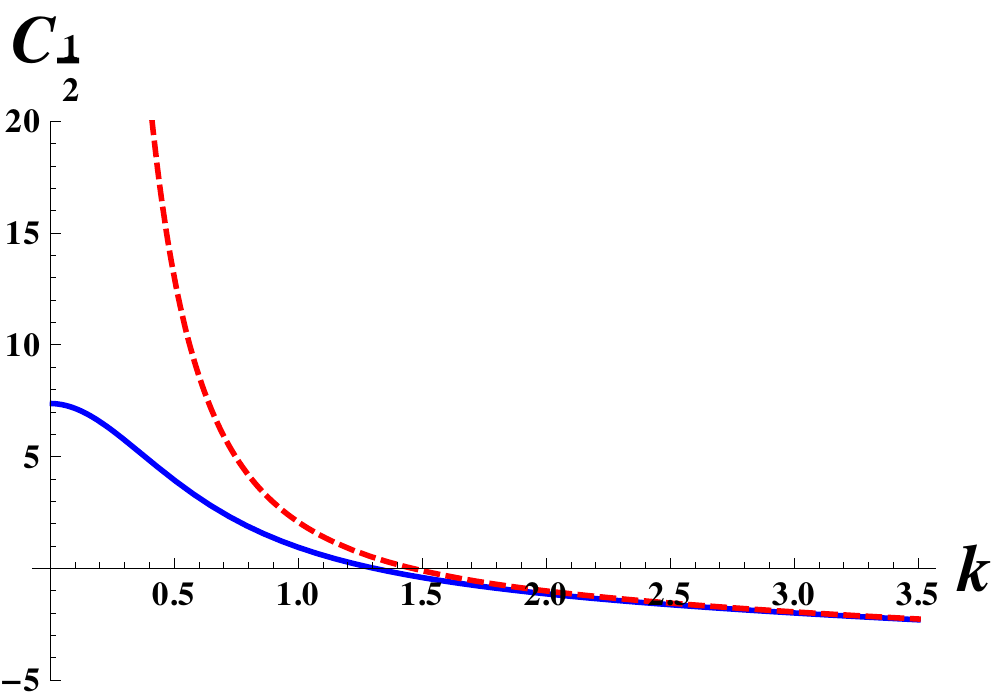}
\caption{\small The red dashed line is $C_{1/2}$ in the background with $\eta=0$, while the blue solid line is the result when a profile for $\eta$ is turned-on.\label{CHalfEta}}
\end{center}
\end{figure}

Plugging our results in the formulas for the holographic correlators (\ref{HolographicC}) and (\ref{HolographicB}), it is easy to see that $C_0$ and $C_1$ are unaffected. On the other hand, both fermionic correlators are modified. As shown in figure \ref{BHalf} the correlator  $B_{1/2}$ has now a non-trivial
dependence on the momenta. Consistently with expectations, it reaches a finite value at zero momentum (hence
providing non-vanishing Majorana mass to SSM gauginos), and falls off
to zero at $k\to\infty$. On the other hand, the pole at $k^2=0$
in $C_{1/2}$ has now disappeared (see figure \ref{CHalfEta}). This is consistent with field theory intuition:  R-symmetry being broken, 't Hooft fermions, if any, cannot couple to
the $j_\alpha$ current and provide zero momentum poles in $C_{1/2}$.
We see the fact that as soon as $\eta$ has a non-trivial profile the correlators $B_{1/2}$ becomes non-vanishing and, {\it at the same time}, the pole in $C_{1/2}$
vanishes, as a remarkable and non-trivial agreement with expectations from the field theory side.

\section{Conclusions and perspectives}
In this paper we have laid out the procedure for computing two-point
functions of operators belonging to a conserved current supermultiplet in a strongly coupled field
theory, using holography. This was aimed, primarily, at
providing a holographic computation for correlators that play a role in models of (general) gauge mediation.

We have been working in the context of 5d consistent truncations of type IIB string theory and
focused our attention on supersymmetry breaking asymptotically AdS
backgrounds. We have found that when R-symmetry is unbroken, the SSM
gauginos generically acquire a Dirac mass by coupling to composite
fermions, which manifest themselves as massless poles in the fermionic
correlator $C_{1/2}$. Sfermions have masses derived from an
integral which converges very nicely in the UV, and are dominated by the pole of the fermionic correlator, providing a spectrum which is reminiscent of gaugino mediation models. On the contrary, for R-symmetry breaking backgrounds the pole in  $C_{1/2}$ disappears, while the R-breaking
correlator $B_{1/2}$ acquires a non-trivial profile, hence providing Majorana mass to SSM gauginos. All our results are in remarkable agreement with field theory expectations \cite{Buican:2009vv,Intriligator:2010be}.

These techniques can be applied, in principle, to several other models and can also find applications in different contexts. Possible further research directions are as follows.

In the present work, we have adopted a top-down approach, and worked in the context of consistent truncations of type IIB 10d string theory.
A different approach is to use a
bottom-up set up. For instance, one could use a model of dynamical SUSY breaking reminiscent
of AdS/QCD set ups. The simplest one is an AdS background with a IR hard
wall\footnote{GGM realizations in warped geometries have already been considered
in \cite{McGarrie:2010yk}. The essential difference between  \cite{McGarrie:2010yk} and our approach is that in the
former, the SUSY-breaking sector is realized as a field theory in the warped geometry,
while in our scenario it is defined by the geometry itself.} (HW). In \cite{hwpaper} we perform an analysis similar to the one we performed in the present work for hard wall backgrounds. The benefit in considering HW setups is that on the one hand
the correlation functions can be established analytically; on the other hand, at the price of loosing any clear string embedding, there is much more
flexibility (for instance, one can allow a profile for $\eta$ providing a VEV for the hidden gauginos bilinear and not a source term).

A natural next step is to try and generalize the present holographic approach
to backgrounds which are not asymptotically
AdS. Indeed, a superconformal theory cannot break supersymmetry spontaneously
(because of the tracelessness of the stress-energy tensor). Thus, the
cases considered here and in \cite{hwpaper}  are nice toy-models but cannot be the full story
for a genuine SUSY breaking hidden sector. It would be interesting, in this respect, to extend our analysis to cascading backgrounds, as those considered in \cite{Benini:2009ff,McGuirk:2009am}.

Eventually, the addition of probe or backreacting D7 branes to
represent the SSM gauge groups will also be a necessary ingredient,
especially if one wants to make contact with the original set up of
holographic gauge mediation
\cite{Benini:2009ff,McGuirk:2009am}. Indeed, in 5d
backgrounds descending from string theory without explicit D-brane sources, the maximal global symmetry
one can have is $SO(6)\simeq SU(4)$, i.e.~the rank is not big enough
to match that of the SSM gauge group. Adding D7 branes is thus
necessary in this top-down approach, even though it makes the
mediation of SUSY breaking less direct and neat.

Finally, let us emphasize that the GGM framework can be seen as an instance of situations in which fields in a visible sector are coupled linearly to composite operators in a strongly coupled hidden sector, and observables are extracted from correlators of such composite operators. This kind of setting is found also in other scenarios, like technicolor-like theories and composite Higgs models, so that one can envisage broader BSM applications along the lines of the holographic calculation we performed here.

\section*{Acknowledgements}
We would like to thank Francesco Bigazzi, Aldo Cotrone, Stefano Cremonesi, Gianguido Dall'Agata, Nick Evans, Anton Faedo, Alberto Mariotti, Carlos Nunez, Gary Shiu and Brian Wecht for useful discussions. We would also like to thank Kostas Skenderis and Marika Taylor for correspondence on \cite{STun}.
The research of R.A. and D.R. is supported in part by IISN-Belgium (conventions
4.4511.06, 4.4505.86 and 4.4514.08) and by a ``Mandat d'Impulsion Scientifique" of the F.R.S.-FNRS. R.A. is a Research Associate of the Fonds de la Recherche Scientifique--F.N.R.S. (Belgium). M.B., L.D.P. and F.P. acknowledge partial financial support by the MIUR-PRIN contract 2009-KHZKRX and by the ESF network ``Holograv".


\end{document}